\documentstyle[12pt]{article}
\begin{document}

{\Large \bf ASHTEKAR CONSTRAINT SURFACE AS\\
PROJECTION OF HILBERT-PALATINI ONE}\\
{\bf V.M.Khatsymovsky}\footnote{Budker Institute of Nuclear
Physics, Novosibirsk 630090, Russia}\\

     The Hilbert-Palatini (HP) Lagrangian of general relativity
being written in terms of selfdual and antiselfdual variables
contains Ashtekar Lagrangian (which governs the dynamics of the
selfdual sector of the theory on condition that the dynamics of
antiselfdual sector is not fixed). We show that nonequivalence
of the Ashtekar and HP quantum theories is due to the specific
form (of the "loose relation" type) of constraints which relate
self- and antiselfdual variables so that the procedure of
(canonical) quantisation of such the theory is noncommutative
with the procedure of excluding antiselfdual variables.

\newpage
1. Ashtekar formulation of general relativity (GR) \cite{Ash}
was shown to follow from Hilbert-Palatini (HP) tetrad-connection
Lagrangian upon partial use of classical equations of motion
together with certain gauge fixing \cite{Rov}. Alternatively,
one can first introduce Ashtekar Lagrangian and show that it
results in HP Lagrangian upon using equations of motion (see,
e.g., reviews \cite{Rom} and, for more detail, \cite{Pel}).
However it may be, equations of motion for selfdual sector from
HP and Ashtekar Lagrangians coincide \cite{AshTat}.

     Another question is that of correspondence between the two
theories, HP and Ashtekar ones, on quantum level. Since the two
Lagrangians differ by more than pure divergence, these theories
may be not equivalent in the framework of canonical
quantisation. Of course, one can present examples in which the
two Lagrangians which do not coincide up to the total divergence
nevertheless provide equivalent theories (the most evident are
examples of locally trivial theories such as 2+1 dimensional
gravity). Therefore it is interesting to study in what cases
such the equivalence takes place and when it does not.

     In this note we show that canonically quantised HP and
Ashtekar theories are inequivalent. To see this it is most
convenient to represent HP theory as a "sum" of two copies of
Ashtekar theory: selfdual and antiselfdual ones, and study
commutators (Dirac brackets) between field variables. The self-
and antiselfdual sectors are not independent: these are related
by some (class II) constraints (which for the pseudoEuclidean
metric signature become well-known reality conditions, but
survive also in the purely real Euclidean case).  Because of the
specific form of class II constraints (which are of the type of
"loose relation" for antiselfdual variables to be excluded) the
commutators of the same variables are different depending on
what theory is quantised, the total HP or only it's Ashtekar
subset. It should be mentioned that this fact is not connected
with the infiniteness of the number of the degrees of freedom of
the theory and may take place even for rather trivial systems,
as it is shown in the simple example below.

     In particular, the quite natural feature of standard GR
that the quantum states cannot describe transition through
unphysical points with degenerate metric shows up in quantum HP
theory as singularity of quantum commutators of field operators
at these points, whereas Ashtekar theory is completely
nonsingular at these points and allows such the transitions.

\bigskip
2. To describe the situation, let us consider the theory whose
phase space is coordinatised by $N$ canonical pairs $(p,q)$ and
$N$ primed canonical pairs $(p^{\prime},q^{\prime})$ subject to
$2N_2$ class II constraints of the form
\begin{equation}
\label{theta-theta}
\Theta_{A}(p,q,p^{\prime},q^{\prime})\stackrel{\rm def}{=}
\theta_{A}(p,q)-\theta_{A}(p^{\prime},q^{\prime})=0           
\end{equation}

\noindent Besides, there are $N_1=N-N_2$ pairs of class I
constraints
\begin{equation}
\Phi_i(p,q)=0,~~~\Phi_i(p^{\prime},q^{\prime})=0              
\end{equation}

     The usual way of quantising the theory with class I
constraints is to impose these constraints on the Hilbert space
of states. As for the class II constraints, these cannot be
imposed on states due to their noncommutativity; the only
consistent way is to take them into account in operator sense by
projecting the fields appearing in Poisson brackets orthogonally
to the surface of class II constraints. Thus we get Dirac
brackets as consistent choice for commutators when performing
canonical quantisation:
\begin{equation}
\{f,g\}_{\rm D}\stackrel{\rm def}{=}\{f,g\}-\{f,\Theta_{A}\}
(\Delta^{-1})^{AB}\{\Theta_{B},g\},                           
\end{equation}

\noindent where $\{f,g\}$ is Poisson bracket and $\Delta^{-1}$
is matrix inversed to that of the Poisson brackets of the class
II constraints:
\begin{equation}
(\Delta^{-1})^{AB}\{\Theta_{B},\Theta_{C}\}=\delta^{A}_{C}
\end{equation}                                                

     Now, if $N_2$, half the number of class II constraints
$\Theta_A$ coincides with the number $N$ of the canonical pairs
$(p^{\prime},q^{\prime})$ to be excluded then the matrix of the
Poisson brackets of $\Theta_A$ is easily invertible:
\begin{equation}
\label{inv-PB-matrix}
\{\Theta ,\Theta\}^{-1}
=\frac{1}{2}\{\theta ,\theta\}_{\eta}^{-1}                    
=\frac{1}{2}\{\eta ,\eta\}_{\theta}
\end{equation}

\noindent Here $\eta=p~~{\rm or}~~q$ and index at brackets means
the set of variables treated as canonical ones w.r.t. which the
brackets are taken. Then simple algebra shows that commutators
coincide with Poisson brackets for the kinetic term $2p\dot{q}$
which should arise in the Lagrangian upon excluding primed
variables (in suggestion that class II constraints are
nondegenerate and give $p^{\prime}=p, q^{\prime}=q$, at least
locally):
\begin{eqnarray}
L & = & 2p\dot{q}-H\\
\{f,g\}_D & = & \frac{1}{2}\{f,g\}_{p,q}\nonumber                      
\end{eqnarray}

\noindent Thus, if the number of (irreducible) class II
constraints equals precisely the number of canonical variables
to be excluded, the latter can be excluded on quantum level.

    If the number of class II constraints is less than the
number of canonical variables to be excluded, we have a freedom
in finding the latter; on the one hand, due to the symmetry
generated by the class I constraints the resulting Lagrangian is
invariant w.r.t. this freedom. On the other hand, the matrix of
the Poisson brackets cannot be inverted in a simple way as in
(\ref{inv-PB-matrix}) and Dirac brackets may be much more
complex as compared to the Poisson ones. For example, consider
the following Lagrangian:
\begin{equation}
L=p_1\dot{q}_1+p_2\dot{q}_2+p^{\prime}_1\dot{q}^{\prime}_1+
p^{\prime}_2\dot{q}^{\prime}_2+
\lambda\Phi(p,q)+\lambda^{\prime}                             
\Phi(p^{\prime},q^{\prime})+\mu_1\Theta_1+\mu_2\Theta_2
\end{equation}

\noindent where $\lambda,~~\lambda^{\prime}$ are Lagrange
multipliers at the first class constraints
\begin{equation}
\label{exmpl-phi}
\Phi(p,q)=p_1-p_2,~~~\Phi(p^{\prime},q^{\prime})
=p^{\prime}_1-                                                
p^{\prime}_2
\end{equation}

\noindent and $\mu_1,~~\mu_2$ are the multipliers at the second
class constraints
\begin{equation}
\label{exmpl-theta}
\Theta_1=p_1+p_2-p^{\prime}_1-p^{\prime}_2,~~
\Theta_2=q_1+q_2-q^{\prime}_1-q^{\prime}_2,~~                 
\end{equation}

\noindent These class II constraints provide the following Dirac
brackets between the coordinates and momenta:
\begin{equation}
\label{Dir-brac-exam}
\{\left (\matrix{p_1\cr
p_2\cr p^{\prime}_1\cr p^{\prime}_2\cr}\right),
\left (\matrix{q_1,q_2,q^{\prime}_1,q^{\prime}_2\cr}         
\right )\}_D=\left (\matrix{
~~\frac{3}{4}~-\frac{1}{4}~~~~\frac{1}{4}~~~~~\frac{1}{4}\cr
-\frac{1}{4}~~~~~\frac{3}{4}~~~~\frac{1}{4}~~~~~\frac{1}{4}\cr
~~~\frac{1}{4}~~~~~\frac{1}{4}~~~~\frac{3}{4}~-\frac{1}{4}\cr
~~~\frac{1}{4}~~~~~\frac{1}{4}~-\frac{1}{4}~~~~\frac{3}{4}\cr}
~\right ).
\end{equation}

\noindent (The brackets of the coordinate-coordinate and
momentum-momentum type are zero).

     Also we can try first to exclude $p^{\prime},~q^{\prime}$
here. The number of constraints $\Theta_1,~\Theta_2$ is
unsufficient to uniquely express $p^{\prime},~q^{\prime}$ in
terms of $p,~q$. However, the class I constraint $\Phi
(p^{\prime},q^{\prime})$ generates the symmetry w.r.t. the
transformation
\begin{equation}
\label{pm-eps}
q^{\prime}_1\rightarrow q^{\prime}_1-\epsilon,
~~~q^{\prime}_2\rightarrow q^{\prime}_2+\epsilon,            
\end{equation}

\noindent which makes it possible to exclude primed variables
from the Lagrangian to give
\begin{equation}
L=2p_1\dot{q}_1+2p_2\dot{q}_2+\lambda (p_1-p_2)              
\end{equation}

\noindent The commutators are defined here, up to an overall
factor, by the Poisson brackets
\begin{equation}
\{\left (\matrix{p_1\cr
p_2\cr}\right ),
\left (\matrix{q_1,q_2\cr}
\right )\}_D=\left (\matrix{                                 
\frac{1}{2}~~~~0\cr
0~~~~\frac{1}{2}\cr}
\right ).
\end{equation}

\noindent This looks simpler than the corresponding upper
lefthand $2\times 2$ block in (\ref{Dir-brac-exam}).

     Qualitatively, we can say that looseness of the constraints
$\Theta_1,~\Theta_2$ in the latter example admits nontrivial
additional degree of freedom (\ref{pm-eps}) which modifies the
commutators.

\bigskip
3. Finally, let us turn to the central point of interest,
namely, Ashtekar formalism considered as reduced HP theory.
Consider this for both cases, namely those of Euclidean and
pseudoEuclidean metric signature (these cases will differ by
minor modifications). The formalism itself is now
well-elaborated (see, e.g., the reviews \cite{Rom}, \cite{Pel}),
and here we only define our notations. The original HP action
\begin{equation}
\label{S}
S=\frac{1}{8}\int \! d^{4} x \: \epsilon_{abcd}
\epsilon^{\mu\nu
\lambda\rho} e^{a}_{\mu} e^{b}_{\nu}
[{\cal D}_{\lambda},{\cal D} _{\rho}]^{cd},
\end{equation}

\noindent where ${\cal D}_{\lambda}=\partial_{\lambda}
+\omega_{\lambda}$ (in fundamental representation) is covariant
derivative, and $\omega_{\mu}^{ab}=-\omega_{\mu}^{ba}$ is
element of $so(3,1)$, Lie algebra of $SO(3,1)$ group in the
pseudoEuclidean case or an element of $so(4)$, Lie algebra of
$SO(4)$ in the Euclidean case. Raising and lowering indices is
performed with the help of metric $\eta_{ab}={\rm diag}(\pm
1,1,1,1)$, while $\epsilon^{0123}=+1$. $\alpha,\beta, \ldots
=1,2,3$ or $\mu,
\nu,\ldots =0,1,2,3$ are coordinate indices and $a,b,
\ldots =0,1,2,3$ or $i,j,\ldots =1,2,3$ are local ones. The
canonical variables are $\omega_{\alpha}$ (coordinates) and
conjugate momenta $\pi^{\alpha}$:
\begin{equation}
\label{pi}
\pi_{ab}^{\alpha}=\frac{1}{2}\epsilon_{abcd}\epsilon^{\alpha
\beta\gamma}e_{\beta}^{c}e_{\gamma}^{d}.                     
\end{equation}

\noindent The antisymmetric tensor fields $A_{ab}$ are split
into selfdual $\,^{+}\!A$ and antiselfdual $\,^{-}\!A$ parts,
which in pseudoEuclidean case take the form
\begin{equation}
A=\,^{+}\!A+\,^{-}\!A,~~~^{\pm}\!A=\frac{1}{2}               
(A\pm i\frac{1}{2}\epsilon^{ab}_{~~cd}A^{cd}).
\end{equation}

\noindent Each tensor part is then embedded into 3D vector space
(complex in pseudoEuclidean case) by expanding over basis of
(anti-)selfdual matrices
\begin{equation}
^{\pm}\!\Sigma^{k}_{ab}=\pm i(\delta^{k}_{a}\delta^{0}_{b}-  
\delta^{k}_{b}\delta^{0}_{a})+\epsilon_{kab},
\end{equation}

\noindent so that
\begin{equation}
^{\pm}\!A^{ab}=\,^{\pm}\!A^{k}\,^{\pm}\!\Sigma^{ab}_{k}/2
\stackrel{\rm def}{=}
\,^{\pm}\!\vec{A}\cdot\,^{\pm}\!\vec{\Sigma}^{ab}/2          
\end{equation}

\noindent In particular, 3-vector images of (anti-)selfdual
constituents of area bivector $\pi^{\alpha}$ in terms of tetrad
take the form
\begin{equation}
\,^{\pm}\!\vec{\pi}^{\alpha}=\frac{1}{2}                     
\epsilon^{\alpha\beta\gamma}
(\mp i\vec{e}_{\beta}\times\vec{e}_{\gamma} -
e^0_{\beta}\vec{e}_{\gamma} + e^0_{\gamma}\vec{e}_{\beta})
\end{equation}

\noindent In the Euclidean case these definitions are modified
by replacing $i=\sqrt{-1}$ by $1$ so that everything becomes
real; besides, we change overall sign of
$\,^{\pm}\!\Sigma^{a}_{kb}$ in order that these would obey
algebra of Pauli matrices times $i$ in both cases. The
(anti-)selfdual projections of area bivector onto the (now real)
3D vector space in terms of tetrad then read
\begin{equation}
\,^{\pm}\!\vec{\pi}^{\alpha}=\frac{1}{2}                     
\epsilon^{\alpha\beta\gamma}
(\mp \vec{e}_{\beta}\times\vec{e}_{\gamma} -
e^0_{\beta}\vec{e}_{\gamma} + e^0_{\gamma}\vec{e}_{\beta})
\end{equation}

The phase space of HP theory can be coordinatised by 9 canonical
pairs $(\,^{+}\!\vec{\pi}^{\alpha},
\,^{+}\!\vec{\omega}_{\alpha})$ and 9 canonical pairs
$(\,^{-}\!\vec{\pi}^{\alpha},\,^{-}\!\vec{\omega}_{\alpha})$.
Each sector of variables, selfdual or antiselfdual ones, is
subject to the same set of class I constraints,
$\Phi_i(\,^{+}\!\pi,\,^{+}\!\omega)$ and
$\Phi_i(\,^{-}\!\pi,\,^{-}\!\omega)$ where 7 functions
$\Phi_i(\pi,\omega)$ take the form
\begin{equation}
\label{ash}
{\cal D}_{\alpha}\vec{\pi}^{\alpha},~~
\epsilon_{\alpha\beta\gamma}\vec{\pi}^{\beta}\cdot           
\vec{R}^{\gamma},~~\epsilon_{\alpha\beta\gamma}
\vec{\pi}^{\alpha}\times\vec{\pi}^{\beta}\cdot
\vec{R}^{\gamma}
\end{equation}

\noindent (Gaussian, diffeomorphism and Hamiltonian constraints,
correspondingly). Here ${\cal
D}_{\alpha}(\cdot)=\partial_{\alpha}
(\cdot)-\vec{\omega}_{\alpha}\times(\cdot)$, $2\vec{R}^{\alpha}=
-\epsilon^{\alpha\beta\gamma}[{\cal D}_{\beta},{\cal
D}_{\gamma}]$, and $(\vec{\pi}^{\alpha},\vec{\omega}_{\alpha})$
are $(\,^{+}\!\vec{\pi}^{\alpha},\,^{+}\!\vec{\omega}_{\alpha})$
or $(\,^{-}\!\vec{\pi}^{\alpha},\,^{-}\!\vec{\omega}_{\alpha})$.
Besides, there are class II constraints ensuring the tetrad form
(\ref{pi}) of area tensors $\pi^{\alpha}_{ab}$,
\begin{equation}
\label{pi-pi}                                                
\,^{+}\!\vec{\pi}^{\alpha}\cdot\,^{+}\!\vec{\pi}^{\beta}-
\,^{-}\!\vec{\pi}^{\alpha}\cdot\,^{-}\!\vec{\pi}^{\beta}=0,
\end{equation}

\noindent and those following by differentiating these in time,
\begin{equation}
\label{pi-pi-D-pi}
\,^{+}\!\vec{\pi}^{\gamma}\cdot\,^{+}\!\vec{\pi}^{(\alpha}   
\times\,^{+}\!{\cal D}_{\gamma}\,^{+}\!\vec{\pi}^{\beta )}-
\,^{-}\!\vec{\pi}^{\gamma}\cdot\,^{-}\!\vec{\pi}^{(\alpha}
\times\,^{-}\!{\cal D}_{\gamma}\,^{-}\!\vec{\pi}^{\beta )}=0.
\end{equation}

\noindent The $(\alpha\dots\beta )$ means the sum of objects
with indices $\alpha\dots\beta$ and $\beta\dots\alpha$. The
(\ref{pi-pi}) and (\ref{pi-pi-D-pi}) relate selfdual and
antiselfdual sectors. Of course, the part of class I constraints
is then the consequence of others. For example, it is sufficient
to set diffeomorphism and Hamiltonian constraints in only one of
the two sectors, selfdual or antiselfdual one, in order that
these would hold modulo other constraints in also another
sector.  This fact is important for establishing the correct
number (two) of the degrees of freedom of gravity system (but it
is unimportant in our analysis).

Thus we arrive at the situation described in section 2 of this
paper where now $(p,q)=(\,^{+}\!\pi,\,^{+}\!\omega )$,
$(p^{\prime},q^{\prime})=(\,^{-}\!\pi,\,^{-}\!\omega )$, the
class II constraints (\ref{theta-theta}) take the form
(\ref{pi-pi}) and (\ref{pi-pi-D-pi}) and class I constraints are
of the type (\ref{ash}). In the simple model given in that
section the class I constraints (\ref{exmpl-phi}) play the role
of Gaussian ones in the problem at hand, while the class II
constraints there (\ref{exmpl-theta}) resemble the recent
(\ref{pi-pi}) and (\ref{pi-pi-D-pi}) (especially this is seen if
one linearizes the latter in the vicinity of the flat
background). Therefore the structure of the Dirac brackets is
similar to that of the simple model; in particular
\begin{equation}
\label{DB-Ash-Ash}
\{\,^{+}\!\pi^{\alpha}_i,\,^{+}\!\omega^k_{\beta}\}_{\rm D}
=\frac{3}{4}\delta^{\alpha}_{\beta}\delta^k_i - \frac{1}{4}
g_{\beta\gamma}\,^{+}\!\pi^{\gamma}_i\frac{1}
{\det\|g_{\alpha\beta}\|}\,^{+}\!\pi^{\alpha k}              
\end{equation}

\noindent and
\begin{equation}
\label{DB-Ash-NonAsh}
\{\,^{+}\!\pi^{\alpha}_i,\,^{-}\!\omega^k_{\beta}\}_{\rm D}
=\frac{1}{4}\delta^{\alpha}_{\beta}\,^{+}\!\pi^{\gamma}_i
\frac{1}{\det\|g_{\alpha\beta}\|}g_{\gamma\epsilon}
\,^{-}\!\pi^{\epsilon k} + \frac{1}{4}g_{\beta\gamma}
\,^{+}\!\pi^{\gamma}_i\frac{1}{\det\|g_{\alpha\beta}\|}
\,^{-}\!\pi^{\alpha k}                                       
\end{equation}

\noindent If, however, we first exclude antiselfdual variables
from the HP Lagrangian using the class II constraints, we find
much more simple expression for nonzero brackets for the rest
variables:
\begin{equation}
\{\,^{+}\!\pi^{\alpha}_i,\,^{+}\!\omega^k_{\beta}\}_{\rm D}=
\{\,^{+}\!\pi^{\alpha}_i,\,^{+}\!\omega^k_{\beta}\}=
\frac{1}{2}\delta^{\alpha}_{\beta}\delta^k_i.                
\end{equation}

Thus, nonequivalence of Ashtekar and HP formulations of gravity
on quantum level has quite transparent (but not quite trivial)
reason and is connected with the "loose relation" type of class
II constraints which allows to express antiselfdual variables in
terms of selfdual ones only up to an SO(3) rotation. In other
words, Ashtekar formulation cannot be obtained from usual tetrad
(or metric) general relativity on quantum level by equivalent
transformations, and this can be explained by quantum
fluctuations of this SO(3) rotation not taken into account in
Ashtekar theory. Important is that this circumstance holds for
both Euclidean and pseudoEuclidean cases. The only difference
between these cases is that in pseudoEuclidean case the (anti-)
selfdual parts become complex, and the class II constraints can
be interpreted as some reality conditions. But the reality of
the Euclidean case does not mean, as we have seen, that problems
connected with class II constraints disappear. In view of this,
the strategy of quantising gravity which consists in considering
the Euclidean Ashtekar theory and then making some generalised
Wick transform to pseudoEuclidean case \cite{Wick} does not
allow to avoid these problems.

If considered from the viewpoint of experimental grounds, the
Ashtekar theory as compared with usual general relativity,
allows metric to pass through the singular point of zero
determinant and thus change the signature. This seems to be far
from reality. In usual HP theory the SO(3) rotation which
connects selfdual and antiselfdual parts strongly fluctuate near
this point thus preventing passing through it.

\end{document}